# Universal moiré nematic phase in twisted graphitic systems

Carmen Rubio-Verdú[1]*, Simon Turkel[1]*, Larry Song[1], Lennart Klebl[2], Rhine Samajdar[3], Mathias S. Scheurer[3,4], Jörn W. F. Venderbos[5,6], Kenji Watanabe[7], Takashi Taniguchi[8], Héctor Ochoa[1], Lede Xian[9], Dante Kennes[2,9], Rafael M. Fernandes[10], Ángel Rubio[9,11,12], Abhay N. Pasupathy[1]

[1]Department of Physics, Columbia University, New York, New York 10027, USA
[2]Institute for Theory of Statistical Physics, RWTH Aachen University, and JARA Fundamentals of Future Information Technology, 52062 Aachen, Germany
[3]Department of Physics, Harvard University, Cambridge, MA 02138, USA
[4]Institute for Theoretical Physics, University of Innsbruck, A-6020 Innsbruck, Austria
[5]Department of Physics, Drexel University, Philadelphia, PA 19104, USA
[6]Department of Materials Science and Engineering, Drexel University, Philadelphia, PA 19104, USA
[7]Research Center for Functional Materials, National Institute for Materials Science, 1-1 Namiki, Tsukuba 305-0044, Japan
[8]International Center for Materials Nanoarchitectonics,
National Institute for Materials Science, 1-1 Namiki, Tsukuba 305-0044, Japan
[9]Max Planck Institute for the Structure and Dynamics of Matter, Center for Free Electron Laser Science, 22761 Hamburg, Germany
[10]School of Physics and Astronomy, University of Minnesota, Minneapolis, 55455 MN
[11]Center for Computational Quantum Physics (CCQ), The Flatiron Institute, 162 Fifth Avenue, New York, NY 10010, USA
[12]Nano-Bio Spectroscopy Group, Departamento de Física de Materiales, UPV/EHU, 20018 Donostia, Spain

* These authors contributed equally to this work.

Abstract: *Graphene moiré superlattices display electronic flat bands. At integer fillings of these flat bands, energy gaps due to strong electron-electron interactions are generally observed. However, the presence of other correlation-driven phases in twisted graphitic systems at non-integer fillings is unclear. Here, we report scanning tunneling microscopy (STM) measurements that reveal the existence of threefold rotational ($C_3$) symmetry breaking in twisted double bilayer graphene (tDBG). Using spectroscopic imaging over large and uniform areas to characterize the direction and degree of $C_3$ symmetry breaking, we find it to be prominent only at energies corresponding to the flat bands and nearly absent in the remote bands. We demonstrate that the $C_3$ symmetry breaking cannot be explained by heterostrain or the displacement field, and is instead a manifestation of an interaction-driven electronic nematic phase, which emerges even away from integer fillings. Comparing our experimental data with a combination of microscopic and phenomenological modeling, we show that the nematic instability is not associated with the local scale of the graphene lattice, but is an emergent phenomenon at the scale of the moiré lattice, pointing to the universal character of this ordered state in flat band moiré materials.*

Clear signatures of correlated electronic phases have been observed in at least three distinct graphene-based moiré systems: magic-angle twisted bilayer graphene (MATBG)[1]–[8], twisted double bilayer graphene (tDBG)[9]–[12] and trilayer rhombohedral graphene on hexagonal boron nitride [13], [14]. In all of these cases, the moiré pattern gives rise to a large unit cell and forms flat bands near the Fermi level. The unambiguously correlated phases that have been observed so far are found at integer fillings of the moiré unit cell. Whether other observed electronic phases are also caused by correlations and how ubiquitous they might be across twisted graphitic systems still remain open questions.

Two of these ordered states — superconductivity and broken lattice rotational symmetry (i.e. nematicity) — have been observed in magic-angle twisted bilayer graphene (MATBG)[4], [5], [8], [15]. Both phenomena are commonly observed in several bulk strongly-correlated materials, most often away from integer fillings[16]. In those cases, experiments that probe the symmetry of Cooper pairs as well as the nematic susceptibility have established beyond doubt that these phases are driven by electron-electron interactions. For graphitic moiré systems, the role of correlations in promoting the superconducting and nematic behaviors remains unclear. In this work, we focus on electron-electron interactions in the nematic phase of graphitic moiré systems.

The nematic phase is a state of matter characterized by a broken rotational symmetry that keeps translational symmetry unaltered[17]. In MATBG, both STM[4], [5], [8] and transport[15] experiments have reported evidence of broken threefold rotational symmetry. However, it is not known whether broken rotational symmetry is a specific property of this system or a more universal feature of twisted graphitic systems in general. Even in MATBG, STM measurements were performed over small areas in samples with significant levels of heterostrain and twist disorder. This makes it difficult to establish whether a true electronic nematic instability exists, or if the observed symmetry breaking over small length scales is due to extrinsic effects unrelated to electron correlations. In bulk materials, electronic nematic phases are often associated with degeneracies related to the orbital and/or spin degrees of freedom. Careful measurements of the properties of the nematic phases in those materials led to insights into the nature of the electronic correlations[18], [19] responsible for them. In twisted graphitic systems, experimental insight into the nature of the emergent interactions that could be responsible for rotational symmetry breaking is still missing.

In this work, we focus on tDBG. Like MATBG, correlated insulating phases have been found in tDBG at integer fillings of the moiré lattice[9]–[12]. Unlike MATBG, where no spin polarization has been reported at half-filling, the insulating phase in tDBG is found to be magnetic. Additionally, the electronic structure of tDBG does not have a magic angle condition, making its electronic properties less sensitive to twist angle and its spatial variations. tDBG therefore provides a more transparent and robust platform than MATBG to investigate the presence of nematic phases, free from extrinsic effects. Studying nematic behavior in tDBG also provides us with a way to explore the relevance of nematicity to the phase diagrams of graphitic moiré systems more broadly.

Our experiments are conducted on moiré regions of tDBG on hexagonal boron nitride using scanning tunnelling microscopy (STM) and spectroscopy (STS) at 5.7 K. Fig. 1a shows an STM topograph over a large area of a tDBG sample. Fig. 1b shows a smaller moiré area, where regions of BAAC stacking (bright spots) are surrounded by inequivalent ABAB (Bernal) and ABCA (rhombohedral) domains[20]. The topograph in Fig. 1a corresponds to a moiré lattice with a twist angle of 1.05 ± 0.02 degrees and a heterostrain of 0.05 ± 0.05 %. Having samples with such twist angle homogeneity and low heterostrain is unprecedented in open-face devices and is crucial to the discussion that follows.

The presence of a large density of states within a narrow energy range of the Fermi level is of central importance to the physics of graphitic moiré systems. Fig. 1c compares STS measurements on each of the three inequivalent high-symmetry stacking configurations at zero doping (i.e. charge neutrality) and zero displacement field. On all three sites, we observe two large peaks in the density of states, one below and one above the Fermi level, corresponding to the valence flat band (VFB) and to the conduction flat band (CFB), respectively. Unlike MATBG, in which the flat bands are restricted to the AA sites, the flat bands in tDBG are spread out with varying spectral weights over the entire moiré unit cell. In addition, we observe higher energy peaks corresponding to the remote valence and conduction bands $RV_2$, $RV_1$, $RC_1$ and $RC_2$ at around -100, -50, +50 and +100 meV, respectively. These remote bands arise from the crossing of the parabolic bands of each Bernal bilayer.

We compare our experimental spectra to theoretical calculations of the local density of states (LDOS) in Figs. 1d and 1e. We consider two complementary models: a tight-binding model (Fig. 1d), which takes into account the material's microscopic nature[21] and an effective low-energy continuum model (Fig. 1e), which is constructed by folding the Dirac dispersions of the four graphene layers[22], [23]. Experimentally, due to the STM geometry, the LDOS is dominated by the top layer. Therefore, to make comparisons with the experimental data meaningful, we project the LDOS onto the top graphene layer, which is where electrons from the STM tip are most likely to tunnel. Additional details can be found in the Supplementary Information (SI). Clearly, despite some minor differences, the STM spectrum at charge neutrality is very well described by both non-interacting models.

In the absence of heterostrain and applied displacement field, the moiré superlattice of tDBG is characterized by a $D_3$ symmetry point group, which contains an out-of-plane threefold rotational axis ($C_3$) and three in-plane twofold rotational axes ($C'_2$). Shown in Fig. 1f are experimental images of the charge neutral and zero field LDOS at energies corresponding to the remote and flat bands shown in Fig. 1c. The VFB and CFB are mostly localized on the ABAB moiré sites. Strikingly complex triskelion structures arise at the energies of the remote bands, which are mostly localized on the ABCA sites. At the energy of the second remote bands and above, the LDOS is peaked on the BAAC sites. The corresponding theoretical LDOS maps obtained from the tight-binding model are shown in Fig. 1g, displaying remarkable agreement with our experimental LDOS maps in Fig. 1f. The fact that all of the structures observed in Fig. 1f preserve threefold rotational symmetry clearly shows that extrinsic effects that could break $C_3$ symmetry such as heterostrain can be neglected in our sample.

Fig. 2a presents gate- and energy-dependent STS maps on 1.05° tDBG obtained in the remote bands as well as in the two flat bands (black dots in Fig. 2d), over gate voltage values ranging from an almost empty valence band (bottom row) to an almost full conduction band (top row). While, for all gate voltages, the LDOS maps of the remote bands are nearly unchanged and retain $C_3$ symmetry, the spatial distribution of the flat band LDOS changes significantly as a function of gate voltage. Most importantly, the VFB LDOS displays pronounced unidirectional stripes at certain values of the gate voltage. This breaking of $C_3$ symmetry over a well-defined range of energy and gate voltage is the main result of the present work.

The application of a back gate voltage in the STM geometry not only induces a nonzero charge doping, but also subjects the sample to a transverse displacement field[20], [24]. Unlike MATBG, the band structure of tDBG is highly sensitive to the presence of an out-of-plane displacement field. Shown in Fig. 2d is the evolution of the spatially averaged (i.e. averaged over positions ABAB, BAAC, and ABCA) spectrum of the sample as a function of gate voltage. The main effect of sample doping is visible as a rigid shift of the spectra along the energy axis. The displacement field, by contrast, directly modifies the band structure and therefore, the density of states. We have calculated the effect of the displacement field on the spatially averaged spectrum within the continuum model in the presence of a self-consistently screened electric field. We find quantitative agreement with experiment (see SI for details). The presence of a displacement field breaks the in-plane $C'_2$ rotational symmetry but leaves the out-of-plane $C_3$ rotational symmetry unaltered, and thus cannot explain the unidirectional stripes observed in experiment.

An LDOS map that respects $C_3$ symmetry will look identical when rotated by 120 degrees. We can therefore quantify the degree of $C_3$ symmetry breaking (i.e. nematicity) by considering differences between a given LDOS map and its $C_3$ rotated counterpart[4]. This quantitative measure of anisotropy is superimposed as a color scale on the site-averaged spectra in Fig. 2d. This clearly reveals that the anisotropy is present at doping values between 0.3 and 0.7 filling of the conduction band and at energies primarily in the VFB. Shown in Fig. 2b is a spectroscopic image of the LDOS in the VFB near half-filling of the CFB, a condition under which the stripes are prominently seen. Overlaid on the image are the moiré lattice high-symmetry site positions. We see that the stripes connect the ABCA and ABAB regions, running in between the darkened BAAC sites and parallel to one of the moiré crystal axes, shown by the dashed lines in Fig. 2b.

We can gain further insight into the symmetry-broken phase by inspecting the site-dependent LDOS. Fig. 2c shows the spectra on each of the three high-symmetry site positions as a function of gate voltage. Here, we draw attention to the spectral shape of the VFB, which is where the stripes are observed. At most gate voltages, the peak's energy corresponding to the location of the VFB (red dots) is identical for the three high-symmetry site positions. However, at the specific gate voltages where the $C_3$ symmetry-broken phase is observed in the LDOS maps, we find that the VFB peaks split, appearing ~4 meV further from the Fermi level on the BAAC site as compared to the ABAB and ABCA sites. While our non-interacting calculations capture the spatially averaged spectrum quite well, they fail to describe this site-dependent splitting of the VFB. The fact that the LDOS develops stripes only at specific values of energy and doping (oriented along a specific crystallographic direction, see below), combined with the inability of single-particle calculations to describe the LDOS spectra at these energies and doping values, is a strong indication that the observed symmetry breaking is driven by electron-electron correlations.

In order to shed light on the phase diagram of tDBG, we can compare our observation of $C_3$ symmetry breaking with the correlated insulator state reported in transport experiments. For the range of twist angles explored in the present work, the equivalent transport experiments report well-developed insulating states for displacement field values between 0.18 and 0.35 V/nm at half-filling of the conduction band and at temperatures below our base temperature[11]. Experimentally, our observations of broken $C_3$ symmetry take place between a range of 0.08 and 0.16 V/nm. We observed similar phenomenology in a second sample at a slightly higher twist angle of 1.15° (see SI). Thus, the existence of a symmetry-broken phase over a broad swath of our parameters space, at values of displacement fields away from the correlated insulator state, indicates that electron-electron interactions are ubiquitous in the phase diagram of tDBG even when insulating phases are absent.

The preceding discussion of $C_3$ symmetry breaking has been restricted to small fields of view around a single moiré unit cell. Since our samples show large and uniform tDBG regions, we can investigate this anisotropy over large length scales, allowing for a direct analysis in Fourier space. In Figs. 3a-b we show the LDOS maps at the energy of the VFB on two overlapping regions of the sample, one near charge neutrality and one near half-filling, respectively. In agreement with what was shown above, nematic behavior is only observed away from charge neutrality. Moreover, the stripe-like pattern extends over hundreds of nanometers and is impervious to the presence of defects, indicating a genuine long-range ordered phase. Shown in the insets of Figs. 3a-b are the corresponding Fast Fourier Transform (FFT) maps. The three moiré Bragg peaks are present at all fillings with varying intensities. No additional peaks beyond these three appear, indicating that translational symmetry is preserved and hence, ruling out a charge-density wave as the origin of the stripe order. Thus, we conclude that a true nematic phase arises in tDBG.

The anisotropy presented in images such as those in Figs. 3a-b can be analyzed in Fourier space by simply considering the intensities of the three moiré Bragg peaks as a function of energy and gate voltage. We show in Fig. 3c-d the evolution of the intensity of these moiré Bragg peaks with energy for each doping condition. Close to charge neutrality, the energy-dependent intensities of the three Bragg peaks remain equal for the whole energy range, as seen in Fig. 3c. However, around half-filling, while the three Bragg peaks have essentially the same intensity across most of the bias range, the peaks split at energies in the VFB, with one of them showing a higher intensity than the other two (see Fig. 3d); in the CFB, they return to equal intensities. The fact that two Bragg peaks in the VFB retain the same intensity while only one of the three is different proves that the nematic director is located along a principal axis of the moiré lattice – otherwise, the three peaks would have three distinct intensities[25].

We can ask whether heterostrain would give rise to similar phenomenology as that observed in the symmetry-broken phase. To this end, we have calculated the LDOS in the continuum model for a range of heterostrain magnitudes ($\varepsilon$) and directions ($\phi$) following Ref. [26]. For each $\varepsilon$ and each $\phi$ we can examine the Fourier peak intensities of the LDOS as a function of energy. Shown in Fig. 3e are the resulting peak intensities for energies in the VFB and CFB. It is clear that for a given $\varepsilon$ and $\phi$, the three Fourier peaks, for most parameters, have three distinct intensities, indicating that symmetry breaking due to heterostrain is, in general, not pointed along a principal moiré axis. In experiment, the observed symmetry breaking in the VFB LDOS is pointed along a principal moiré axis. Moreover, the dependence of the peak intensity shows nearly identical contours for both the CFB and VFB LDOS. Strain-induced symmetry breaking is therefore expected to produce similar signatures in Fourier space over wide energy ranges spanning both flat bands. In experiment, we observe $C_3$ broken Fourier peaks intensities only in the VFB and only at specific gate voltages. Fig. 3f shows the continuum model's LDOS in each flat band in the presence of heterostrain. Neither bears resemblance to the experimental images.

Having ruled out heterostrain and displacement field as possible external causes for $C_3$ symmetry breaking, we investigate theoretically whether interactions could promote an electronic nematic state that spontaneously breaks $C_3$ symmetry. While previous theoretical works have discussed nematicity in MATBG[25], [27]–[35], very few deal with $C_3$ symmetry breaking in tDBG[27]. Starting from our microscopic tight-binding model at half-filling of the CFB, we add a screened Coulomb repulsion that is cut off at a distance of ~3.7 Å (Fig. 4a). Using an unbiased, beyond mean-field functional renormalization group (fRG) approach[36], we find several closely competing leading instabilities in the charge channel (details on the tight-binding model, interactions and fRG are in the SI). Importantly, one of the leading instabilities (Fig. 4b) spontaneously breaks threefold rotational symmetry but not translational symmetry, i.e. it is a nematic phase on the moiré scale. It is manifested as pronounced threefold anisotropic features in the LDOS for low energies (Fig. 4b), in agreement with the STS images for the same doping levels.

While the fRG calculations prove the existence of a strong effective interaction in the nematic channel in tDBG, the limitations imposed by their high computational costs make it challenging to systematically study the spatial and energetic dependence of the nematic order parameter. Therefore, and to gain a better understanding of the underlying physics, we move to the continuum model and write down the most general form of the continuum nematic order parameter in layer, sublattice, valley, and spin space (see details in the SI). Given the large number of quantum numbers, one can write many different nematic order parameters that break the $C_3$ symmetry of the moiré lattice. Although symmetry enforces all order parameters to be nonzero in the nematic phase, it is interesting to ask whether the experimentally observed features in the LDOS point to a dominant one. Specifically, we consider two opposite limits: a nematic order parameter that breaks $C_3$ at the scale of the atomic lattice (dubbed *graphene nematic* and illustrated schematically in Fig. 4c), and a nematic order parameter that breaks $C_3$ symmetry only at the scale of the moiré lattice by making its bonds inequivalent (dubbed *moiré nematic*, see Fig. 4d). While graphene nematicity is only sensitive to the existence of the moiré lattice because of the imposed interlayer potential, moiré nematicity is insensitive to the local properties of the layers.

In Fig. 4e, we show the resulting LDOS at the three high-symmetry site positions of tDBG (ABAB, BAAC, ABCA) for the non-interacting, moiré nematic, and graphene nematic cases. Clearly, only in the case of moiré nematicity does the VFB peak at position BAAC split from the peaks at the other two positions similar to what is observed experimentally in Fig. 2c and also reproduced by the fully microscopic model. The real-space images of the LDOS at energies corresponding to the remote and flat bands, shown in Figs. 4f-h, further reveal that in the case of moiré nematicity, the threefold anisotropy is much more pronounced in the VFB, is smaller but noticeable in the CFB, and is negligible in the remote bands. In contrast, the changes in the LDOS are almost imperceptible in the case of graphene nematicity. The combination of the results from the tight-binding and continuum models suggests that the nematicity observed in tDBG is an emergent instability of the moiré superlattice.

Our experimental study reveals that nematicity is an integral part of the phase diagram of tDBG, being realized in a wide doping range away from charge neutrality. A careful comparison of the experimental phenomenology of the nematic state with theory points to an emergent nematic electron-electron interaction that operates at the scale of the moiré period, rather than at the scale of the local graphene lattice. This, together with previous manifestations of the nematic phase in MATBG, indicate that the nematic phase is a ubiquitous property of moiré systems with flat bands, and is not specific to the details at the atomic scale. One might therefore expect threefold rotational symmetry breaking to be also present in carbon-free flat band moiré lattices, a subject for future experimental investigation.

**Methods**

We fabricated tDBG samples following the tear-and-stack method. PPC was used as a polymer to pick up hBN, then half of a piece of bilayer graphene, followed by the second half twisted relative to the first half. Then we flipped over the structure and placed it on top of a Si/SiO2 chip. We made direct contact to the tDBG structure via microsoldering with Field's metal[37].

Ultra high-vacuum Scanning Tunneling Microscopy and Spectrocopy were carried out in a home-built STM at 5 K. The tips were prepared on clean Au(111) surface and calibrated to be atomically sharp and to detect the Au(111) Shockley surface state via STS.

**Acknowledgements**

This work was supported by Programmable Quantum Materials, an Energy Frontier Research Center funded by the U.S. Department of Energy (DOE), Office of Science, Basic Energy Sciences (BES), under award DE-SC0019443. STM equipment support was provided by the Air Force Office of Scientific Research via grant FA9550-16-1-0601. C.R.V. acknowledges funding from the European Union Horizon 2020 research and innovation programme under the Marie *Skłodowska*-Curie grant agreement No 844271. This work is supported by the European Research Council (ERC-2015-AdG-694097), Grupos Consolidados (IT1249-19) and the Flatiron Institute, a division of the Simons Foundation. We acknowledge funding by the Deutsche Forschungsgemeinschaft (DFG) under Germany's Excellence Strategy - Cluster of Excellence Matter and Light for Quantum Computing (ML4Q) EXC 2004/1 - 390534769 and Advanced Imaging of Matter (AIM) EXC 2056 - 390715994 and funding by the Deutsche Forschungsgemeinschaft (DFG) under RTG 1995 and GRK 2247. Support by the Max Planck Institute - New York City Center for Non-Equilibrium Quantum Phenomena is acknowledged. H.O. is supported by the NSF MRSEC program grant No. DMR-1420634. Tight-binding and fRG simulations were performed with computing resources granted by RWTH Aachen University under projects rwth0496 and rwth0589. R.S. and M.S. acknowledge support from the National Science Foundation under Grant No. DMR-2002850. RMF was supported by the DOE-BES under Award No. DE-SC0020045. K.W. and T.T. acknowledge support from the Elemental Strategy Initiative conducted by the MEXT, Japan ,Grant Number JPMXP0112101001,  JSPS KAKENHI Grant Number JP20H00354 and the CREST(JPMJCR15F3), JST.

**Author contributions**

C.R.V. and S.T. performed the STM measurements. L.S. fabricated the samples for STM measurements. C.R.V. and S.T. performed experimental data analysis. K.W. and T.T. provided hBN crystals. L.K. L.X. and D.M.K. performed tight-binding calculations. S.T., R.S., M.S.S., J.W.F.V., H.O. and R.M.F. performed continuum-model calculations. R.M.F., A.R. and A.N.P. advised. C.R.V. and S.T. wrote the manuscript with assistance from all authors.

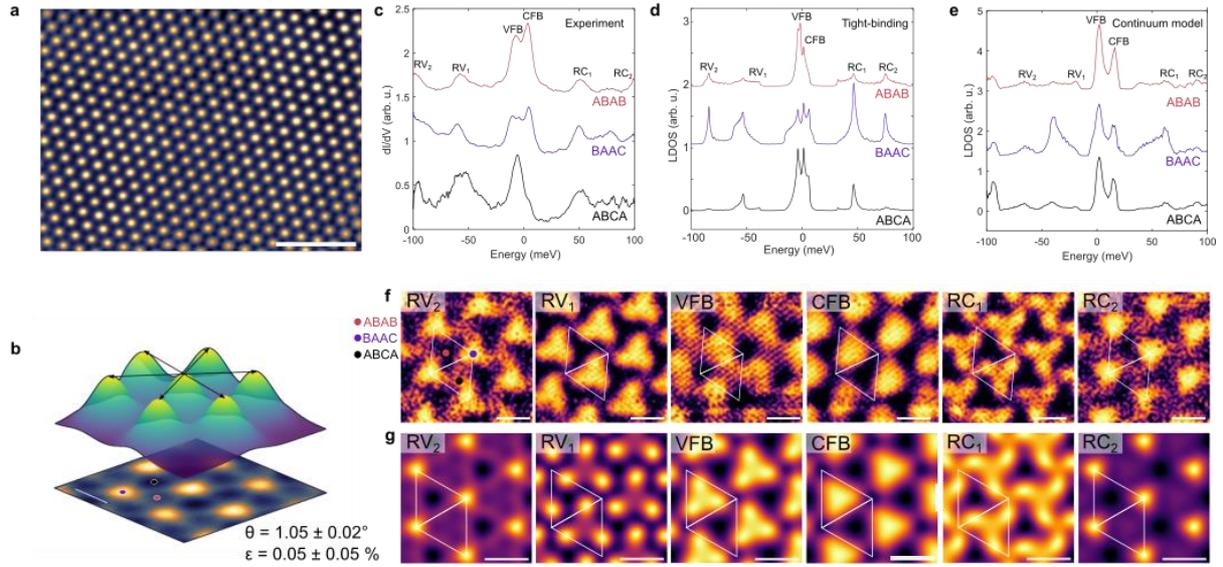

**Fig. 1. LDOS in twisted double bilayer graphene. a**, STM topography on 1.05° tDBG (scale bar represents 100nm). **b**, Zoomed-in image of an STM topograph and three-dimensional representation of the method used to obtain the heterostrain values. Following Ref.[4], we fitted a Gaussian function to the maximum of each BAAC site, thus obtaining the three moiré wavelengths as indicated with the black arrows. **c**, dI/dV at zero gate voltage on BAAC, ABCA and ABAB sites in 1.05° tDBG. **d-e**, Tight-binding and continuum model calculations of the LDOS on BAAC, ABCA and ABAB sites. **f**, LDOS map at the energies indicated in panel c at zero doping. The three inequivalent sites are indicated by the circles. **g**, Tight-binding simulations of the LDOS at the energies indicated in d. The moiré unit cell is indicated in white. Scale bars in b, f, g correspond to 10 nm. Curves are offset for clarity.

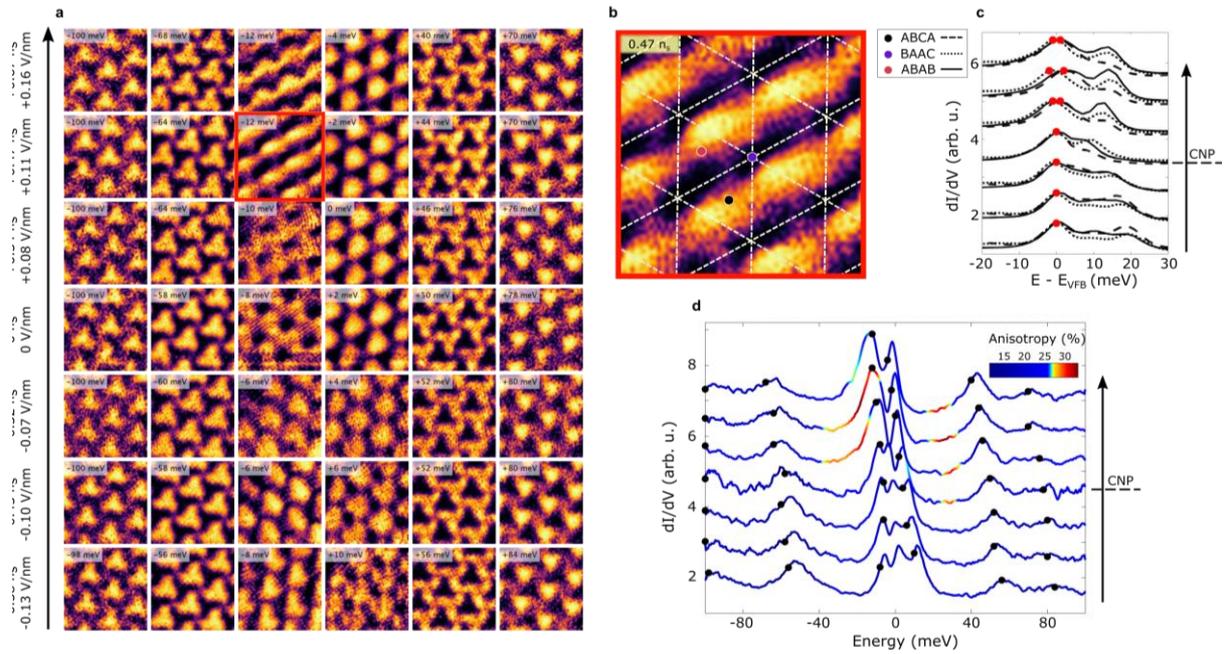

**Fig. 2. Broken C$_3$ symmetry. a,** dI/dV maps at the energy, doping and displacement field conditions indicated. **b**, LDOS map at the energy of the VFB and at the gate voltage in which broken rotational symmetry is observed. The three inequivalent lattice sites are shown as well as the moiré unit cells (dashed lines). **c**, dI/dV spectra on the three inequivalent sites at the doping conditions shown in a. The red dots indicate the energy position of the valence flat band on the different sites. **d,** dI/dV averaged over BAAC, ABAB and ABCA sites at the doping conditions shown in a. The color scale represents the anisotropy values obtained following Ref.[4]. Curves are offset for clarity.

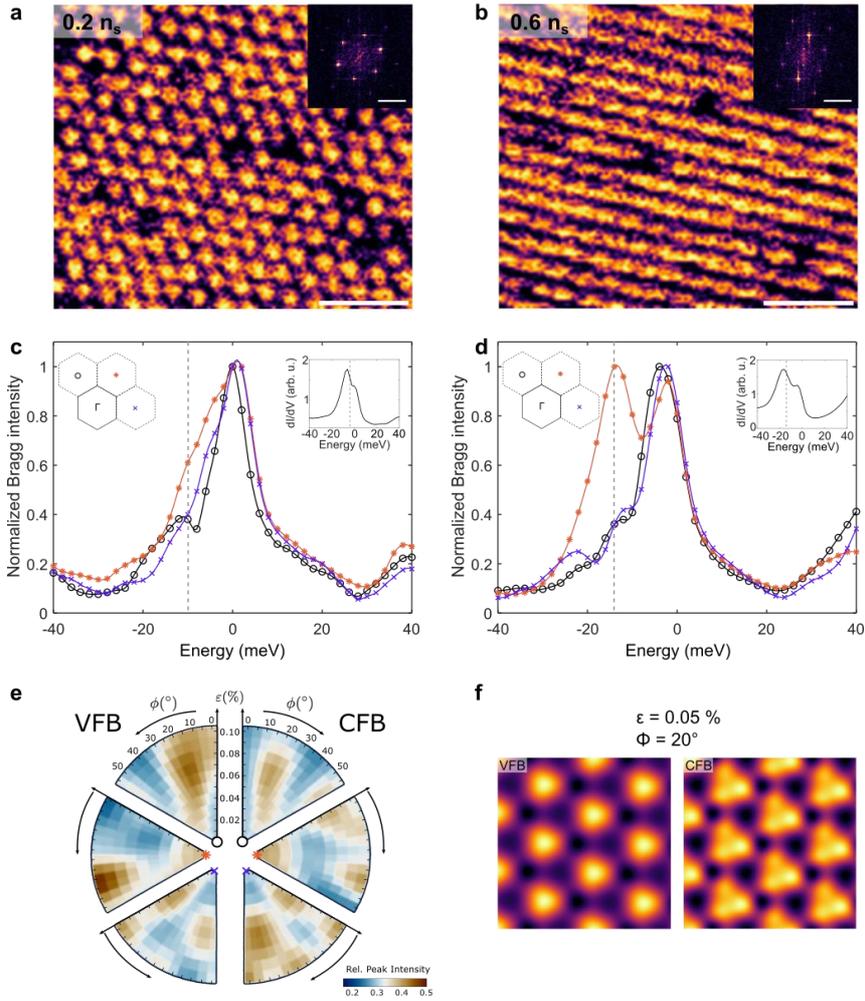

**Fig. 3. Manifestations of long-range nematic order in tDBG. a**, dI/dV maps at the energy of the valence flat band **a**, close to charge neutrality and **b**, around half-filling of the conduction flat band. The insets show the FFT of each LDOS map. The scale bar corresponds to 125 nm **c-d**, Energy evolution of the normalized intensity of the three moiré Bragg peaks. The first and second Brillouin zones are shown in the inset as well as the average spectrum over the regions shown in a and b. Dashed lines indicate the energy of the maps shown in a and b. **e**, Bragg peaks' intensity from the continuum model LDOS maps at the valence and conduction flat bands' energy. Each of the six panels maps the evolution of the Bragg peak with heterostrain strength and strain angle, which is defined with respect to the horizontal of the LDOS maps. **f**, Continuum model LDOS maps at the energy of the valence and conduction flat bands under the presence of 0.05 % strain at 20° strain angle.

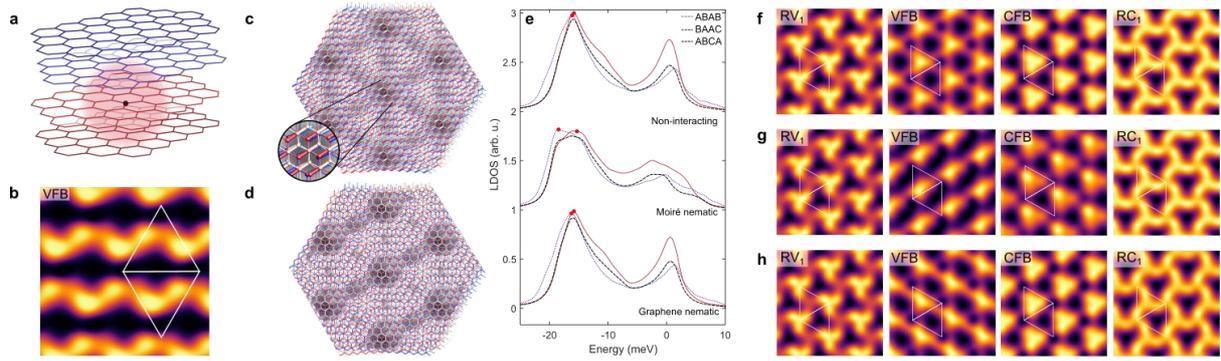

**Fig. 4. Moiré nematic order**. **a**, tDBG schematic. The red sphere represents the range of interactions included in the tight-binding calculation. **b**, Tight-binding LDOS map corresponding to the competing instability that spontaneously breaks rotational symmetry. **c-d**, Twisted double bilayer graphene moiré patterns under the presence of graphene and moiré nematic order parameters, respectively. **e**, Non-interacting, moiré nematic and graphene nematic continuum model LDOS. The red dots indicate the energy position of the valence flat band. **f-h**, Unperturbed, moiré nematic and graphene nematic continuum model LDOS maps, respectively. The moiré unit cell is shown in white.

# Supplementary Information

1. **Continuum model LDOS under an external displacement field.**

As described in the main text, the back gate voltage in the STM geometry induces charge doping as well as an external displacement field perpendicular to the graphene sample. In order to obtain the experimental displacement field values, we performed continuum model simulations. We calculated spatially-averaged LDOS in the presence of a self-consistently screened electric field. The results are shown in Supplementary Figure 1. On the left panel we show experimental LDOS at a wide range of back gate voltage values. We found that the experimental spectra are better reproduced with displacement field values within ±0.2 V/nm, as shown in the right panel of Supplementary Figure 1.

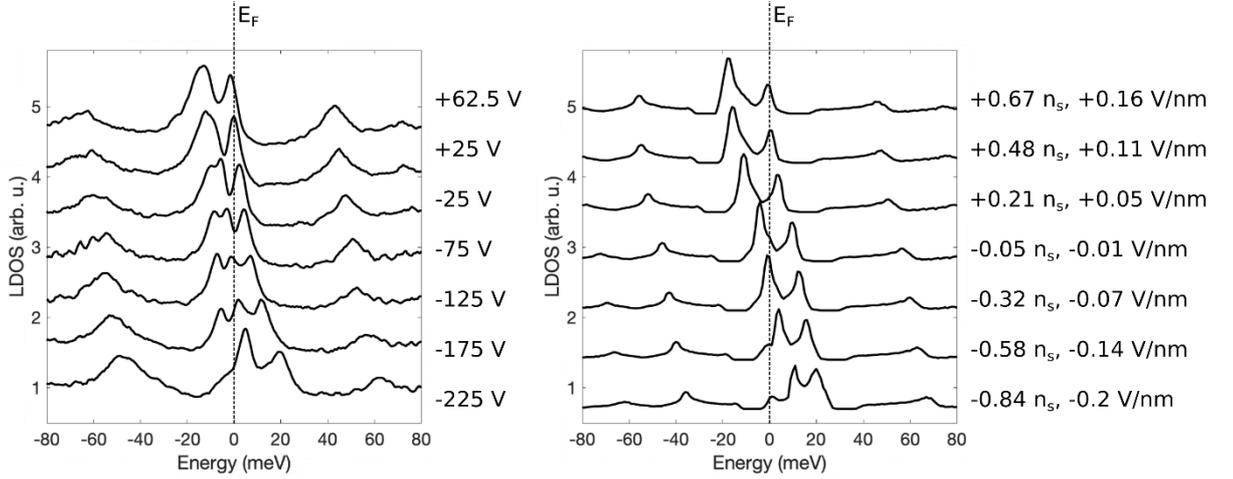

**Supplementary Figure 1. 1.05° tDBG under the presence of a displacement field**. (left) Experimental LDOS on tDBG at several back-gate values. (right) Continuum model LDOS under the presence of a displacement field. The doping and displacement field values are included. The vertical dashed lines point the Fermi level.

2. **tDBG at 1.15° twist angle.**

We performed additional low-temperature STM experiments on a second tDBG sample at a twist angle of 1.15°. We found similar phenomenology for both 1.15° and 1.05° twisted double bilayer graphene, namely broken rotational symmetry at the energy of the valence flat band around the half-filling condition. Supplementary Figure 1a shows a large, uniform region of 1.15° tDBG. As described in the main text, stripe order emerges at the energy of the valence flat band at around half-filling, as shown in Fig. S1b. Rotational symmetry is recovered when moving back to the charge neutrality point (see Fig. S1c).

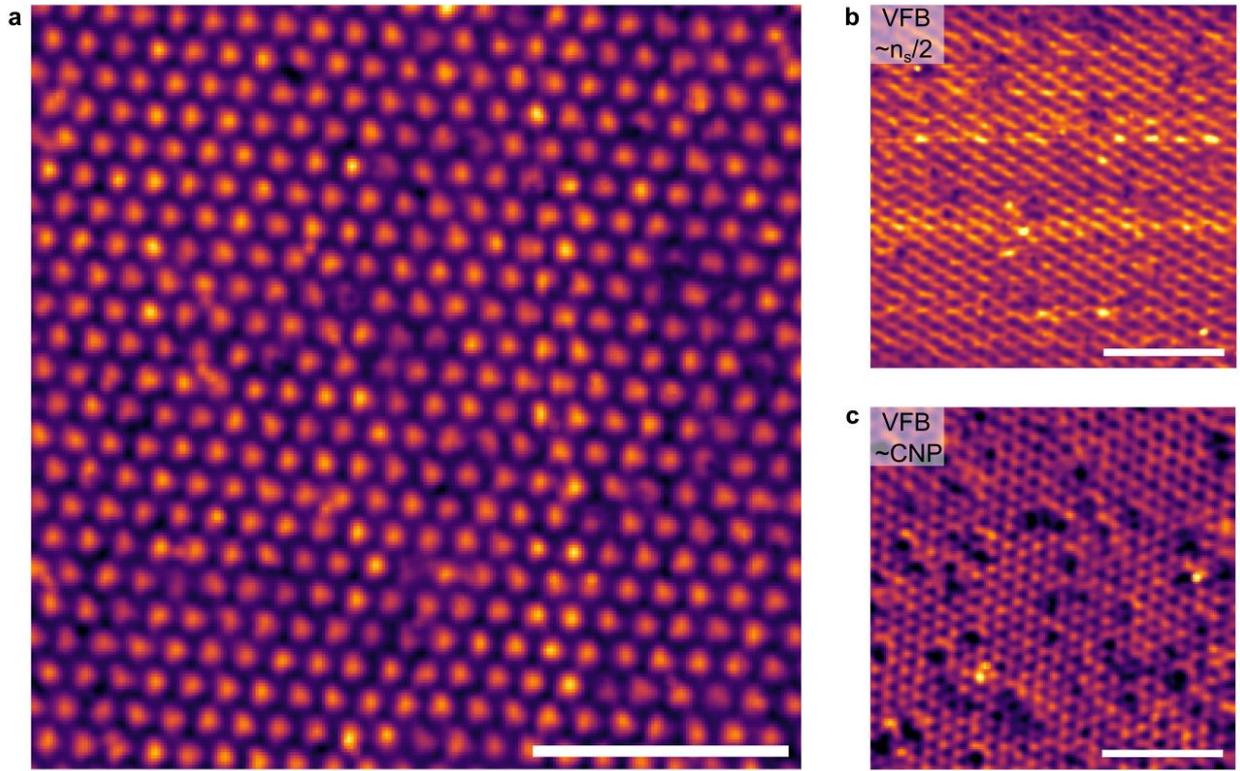

**Supplementary Figure 2. Twisted double bilayer graphene at 1.15° twist angle**. **a**, LDOS map on a 300x300nm² region of uniform moiré. **b**, **c**, LDOS map on the same region as a, at the energy of the valence flat band and at half-filling and charge neutrality point, respectively. Scale bars correspond to 100 nm.

3. **Tight Binding Model.**

The band structure of tDBG is microscopically generated from a tight-binding model as presented in Ref. [S1]. We use a commensurate atomic structure with a twist angle of $\theta \approx 1.1°$ that has 2232 sites per unit cell. The atomic structure is relaxed using the LAMMPS code [S2] with the same parameters as described in Ref. [S3]. The intralayer interactions within each graphene layer are modeled via the second-generation reactive empirical bondorder (REBO) potential [S4]. The interlayer interactions are modeled via the Kolmogorov-Crespi (KC) potential [S5], using the recent parametrization of Ref. [S6]. The relaxation is performed using the fast inertial relaxation engine (FIRE) algorithm [S7]. We set the hopping parameters to $V_{pp\pi}^0 = -3.24$ eV and $V_{pp\sigma}^0 = 0.48$ eV which reproduces our experimental findings of the LDOS well. The intrinsic symmetric polarization energy (onsite potentials for the inner layers) is set to -32.76 meV. The non-interacting part of the Hamiltonian then reads

$$\mathcal{H}_0 = \sum_{i,j,\sigma} V_{ij} c^\dagger_{i,\sigma} c_{j,\sigma}$$

where $V_{ij}$ denotes the resulting hopping amplitude between sites $i$ and $j$ and $c^{(\dagger)}_{i,\sigma}$ annihilates (creates) an electron on site $i$ with spin $\sigma$.

We use an Ohno type interaction [S8] that is screened for $r = 0$ and truncated by a sharp cutoff function at $r_c = 1.5a = 3.69$ Å. The interacting part of the Hamiltonian reads

$$\mathcal{H}_{\text{int}} = \sum_{i,j,\sigma,\sigma'} \frac{U\Theta(r_c - |r_i - r_j|)}{\sqrt{1 + \left((r_i - r_j)/r_s\right)^2}} \rho_{i,\sigma} \rho_{j,\sigma'}$$

with the density-operator $\rho_{i,\sigma} = c^\dagger_{i,\sigma} c_{i,\sigma}$ and $r_s = 3a = 7.38$Å. For our fRG simulations, we set the interaction strength to $U = 6$ eV and $U = 8$ eV.

### 4. Functional Renormalization Group

The fRG treats interactions perturbatively in an unbiased way beyond mean-field by interpolating from the free system at high energies to an effective low-energy model. We use the truncated unity, intraorbital bilinear and Γ-point approximations from Ref. [S9]. Within these approximations, the vertex function $V^\Lambda_{o_1 o_2 o_3 o_4}$ is split up into three channels (pairing, crossed particle-hole, direct particle-hole) that are matrices in unit cell index space:

$$P^\Lambda_{o_1 o_2 o_3 o_4} = \delta_{o_1 o_2} \delta_{o_3 o_4} P^\Lambda_{o_1 o_3}$$
$$C^\Lambda_{o_1 o_2 o_3 o_4} = \delta_{o_1 o_4} \delta_{o_2 o_3} C^\Lambda_{o_1 o_3}$$
$$D^\Lambda_{o_1 o_2 o_3 o_4} = \delta_{o_1 o_3} \delta_{o_2 o_4} D^\Lambda_{o_1 o_2}$$

with

$$V^\Lambda_{o_1 o_2 o_3 o_4} = P^\Lambda_{o_1 o_2 o_3 o_4} + C^\Lambda_{o_1 o_2 o_3 o_4} + D^\Lambda_{o_1 o_2 o_3 o_4}$$

The differential equations in the energy cutoff parameter $\Lambda$, which describe the renormalization group flow of a spin-degenerate system, read

$$\frac{d}{d\Lambda} \hat{P}^\Lambda = \hat{V}^{\mathbb{P}_P,\Lambda} \dot{\hat{L}}^{\text{PP},\Lambda} \hat{V}^{\mathbb{P}_P,\Lambda}$$
$$\frac{d}{d\Lambda} \hat{C}^\Lambda = \hat{V}^{\mathbb{P}_C,\Lambda} \dot{\hat{L}}^{\text{PH},\Lambda} \hat{V}^{\mathbb{P}_C,\Lambda}$$
$$\frac{d}{d\Lambda} \hat{D}^\Lambda = -2\hat{V}^{\mathbb{P}_D,\Lambda} \dot{\hat{L}}^{\text{PH},\Lambda} \hat{V}^{\mathbb{P}_D,\Lambda}$$

$$+\hat{V}^{\mathbb{P}_D,\Lambda}\dot{\hat{L}}^{\text{PH},\Lambda}\hat{V}^{\mathbb{P}_C,\Lambda} + \hat{V}^{\mathbb{P}_C,\Lambda}\dot{\hat{L}}^{\text{PH},\Lambda}\hat{V}^{\mathbb{P}_D,\Lambda}$$

All quantities are connected by matrix products. The channel projections of the vertex function are given by

$$\hat{V}^{\mathbb{P}_P,\Lambda} = \hat{P}^\Lambda + \text{diag}(\hat{C}^\Lambda) + \text{diag}(\hat{D}^\Lambda)$$
$$\hat{V}^{\mathbb{P}_C,\Lambda} = \text{diag}(\hat{P}^\Lambda) + \hat{C}^\Lambda + \text{diag}(\hat{D}^\Lambda)$$
$$\hat{V}^{\mathbb{P}_D,\Lambda} = \text{diag}(\hat{P}^\Lambda) + \text{diag}(\hat{C}^\Lambda) + \hat{D}^\Lambda$$

The particle-particle $(\hat{L}^{\text{PP},\Lambda})$ and particle-hole $(\hat{L}^{\text{PH},\Lambda})$ loops within these approximations read

$$\hat{L}^{\text{PP},\Lambda} = \int \frac{dk_0}{2\pi} \frac{1}{N} \sum_{\vec{k}} \hat{G}^\Lambda(k) \circ \hat{G}^\Lambda(-k)$$
$$\hat{L}^{\text{PH},\Lambda} = \int \frac{dk_0}{2\pi} \frac{1}{N} \sum_{\vec{k}} \hat{G}^\Lambda(k) \circ \hat{G}^\Lambda(k)$$

with $\hat{A} \circ \hat{B}$ the element-wise matrix product. For our fRG simulations, we use a sharp frequency cutoff of the Green's function that allows us to trivially carry out the Matsubara frequency integrals in the above equations:

$$\hat{G}^\Lambda(k) = \hat{G}^{(0)}(k)\sqrt{\Theta(|k_0| - \Lambda)}$$

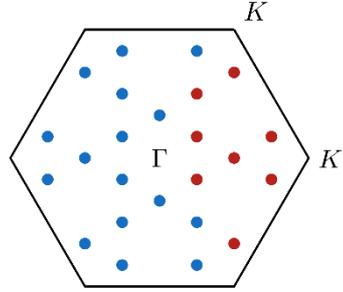

**Supplementary Figure 3.** Momentum mesh used in fRG simulations. Red points: meshing of irreducible Brillouin zone (IBZ), blue points and red points: meshing of full Brillouin zone. In order to reduce the computational effort required to obtain the particle-particle and particle-hole loops $(\hat{L}^{\text{PP},\Lambda}$ and $\hat{L}^{\text{PH},\Lambda})$, we use the IBZ meshing and reconstruct the full loops using the in-plane $C_3$ symmetry.

The meshing of the momentum part of the integrals contains N = 24 points in the moiré Brillouin zone (see Supp. Fig. 3). We start the fRG flow at an energy cutoff of $\Lambda = 10$ eV and write the initial interaction in the direct particle-hole ($D$) channel. The chemical potential $\mu = \mu_n + 3.83$ meV corresponds to a partially filled conduction band, where $\mu_n$ is the chemical potential at charge neutrality. During each step of the fRG flow, the leading eigenvalue of each interaction channel is determined and the stopping condition whether one of the eigenvalues is larger than $3 \cdot 10^2$ eV is evaluated. At the end of the flow, an eigendecomposition of the leading channel reveals the order parameters associated with the phase the system will likely order in. Mean-field decoupling of the bilinears associated with the divergent channel enables us to determine single-particle properties in the ordered phase. The parameters presented here lead to a divergence in the $D$ channel that points to a charge-modulation instability. We perform a mean-field decoupling of the (spin-independent) order parameter and arrive at the effective Hamiltonian (neglecting constants)

$$\mathcal{H}_{\text{eff}} = \mathcal{H}_0 + U_{\text{eff}} \sum_{k,i,\sigma} D_i^\alpha c_{k,i,\sigma}^\dagger c_{k,i,\sigma}$$

The effective coupling strength $U_{\text{eff}}$ remains unknown from our simulations, we set the *ad-hoc* value of $U_{\text{eff}} = U$. The order parameter $\vec{D}^\alpha$ is the $\alpha$-th eigenvector of the effective direct particle-hole channel and describes charge redistribution within the moiré unit cell.

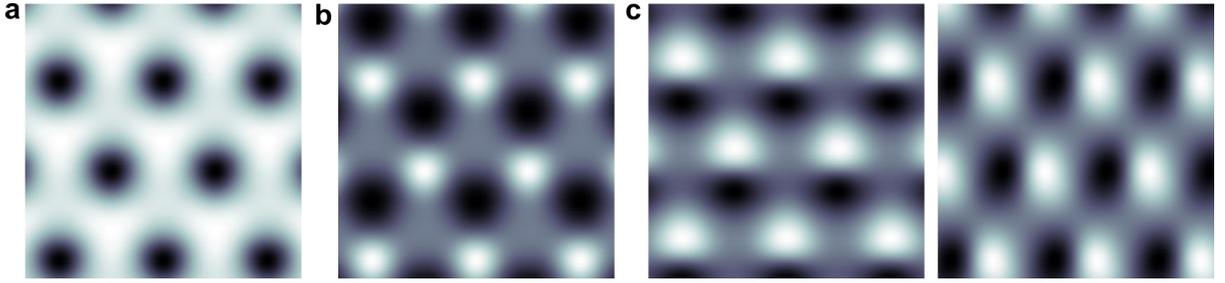

**Supplementary Figure 4.** Top layer projection of the three leading charge modulation orderings (leading eigenvectors) from our simulations. The first two orderings **a**, **b**, respect the lattice symmetries and manifest themselves as charge reordering within the moiré unit cell. The third ordering breaks the in-plane threefold rotational symmetry and consists of two degenerate eigenvectors shown in panel **c**. As the variation of the order parameter is on the moiré scale, it describes *moiré nematicity.*

The first two eigenvectors (orderings) respect the system's lattice symmetries whereas the third and fourth eigenvectors (degenerate eigenvalues) break C3 (see Supp. Fig. 4). We note that in the fRG flow there are multiple closely competing orderings. The second and third orderings are weaker only by factors of 0.85 and 0.83. Therefore, all three leading ordering tendencies may strongly influence the physics at play and be visible experimentally. In the case of a degenerate eigenvalue (i.e. due to breaking a lattice symmetry), any linear combination of the two eigenvectors can be the order parameter. Due to the approximations made, it is possible that the closely competing orderings are interchanged making e.g. moiré nematicity the leading instability. For the LDOS obtained from the effective Hamiltonian (shown in Fig. 4b in the main text), we use the first of the two degenerate eigenvectors (corresponding to the third eigenvalue) as order parameter.

### 5. Nematic instability within the continuum model.

The band structure of tDBG is well-described by the continuum model proposed in Ref. [S10]. We set the twist angle to be 1.05° and use the same model parameters as Ref. [S10] with two exceptions. The first difference is that we take a rescaled band velocity of $\hbar v/a = 2.776$ eV. Secondly, instead of assuming a constant value for the interlayer asymmetric potential induced by the gate electric field, we use on-site potentials $\Delta_1 = 4.079$ meV, $\Delta_2 = 1.021$ meV, $\Delta_3 = -1.537$ meV, and $\Delta_4 = -3.563$ meV for layers 1 through 4, respectively. This is done to self-consistently include the effect of the displacement field.

The nematic order can be described by a two-component order parameter $\boldsymbol{\Phi} = \Phi_0(\cos 2\theta, \sin 2\theta)$, with the angle $\theta$ characterizing the orientation of the nematic director [S11]. The electronic degrees of freedom are described by the field operator $a_{\sigma,\ell,s,\tau}(\boldsymbol{r})$, with spin $\sigma = \uparrow, \downarrow$, sublattice s = 1,2, layer $\ell$ = 1, 2, 3, 4, valley $\tau = \pm$, and (continuum) position $\boldsymbol{r} \in \mathbb{R}^2$. While there are many ways of writing down the coupling between $\boldsymbol{\Phi}$ and the electronic degrees of freedom in the continuum model, we here focus on two opposing limiting cases, which we dub moiré nematic and graphene nematic.

The moiré nematic order parameter should be thought of as an anisotropic deformation of the effective nearest-neighbor hopping amplitudes on the moiré scale. Its coupling to the electrons is written as:

$$\mathcal{H}_{\boldsymbol{\Phi}} = \sum_{\boldsymbol{R}} \boldsymbol{\Phi} \cdot \boldsymbol{\phi}(\boldsymbol{r}, \boldsymbol{R}) \, a^\dagger_{\sigma,\ell,s,\tau}(\boldsymbol{r}+\boldsymbol{R}) a_{\sigma,\ell,s,\tau}(\boldsymbol{r}) + \text{H.c.}$$

where $\boldsymbol{R}$ denotes moiré lattice vectors. Restricting $\boldsymbol{R}$ to nearest neighbors and neglecting the $\boldsymbol{r}$ dependence of $\boldsymbol{\phi}$, this corresponds to a momentum-dependent shift of the spectrum $E_{n\boldsymbol{k}}$:

$$E_{n\boldsymbol{k}} \to E_{n\boldsymbol{k}} + \boldsymbol{\Phi} \cdot \boldsymbol{f}(\boldsymbol{k}), \quad \boldsymbol{f}(\boldsymbol{k}) = \frac{8}{3}\left(\cos k_y - \cos\frac{\sqrt{3}k_x}{2}\cos\frac{k_y}{2}, \sqrt{3}\sin\frac{\sqrt{3}k_x}{2}\sin\frac{k_y}{2}\right)^T$$

Here, $\boldsymbol{k}$ denotes the momentum in the first moiré Brillouin zone, and $n$ labels the bands of the continuum model without nematic order. As shown in Fig. 4 of the main text, upon setting $\Phi_0 = 1$ meV, $\theta = \pi$, this model not only reproduces the experimentally observed stripes in the LDOS images, but it also captures the splitting between the valence-flat-band peaks of the d$I$/d$V$ curves at the ABAB, BAAC, and ABCA site positions.

In the case of the graphene nematic order parameter, nematicity occurs on the scale of the underlying graphene lattice. The simplest form of its coupling to the electrons is given by:

$$\mathcal{H}_\Phi = \int d\boldsymbol{r}\, \boldsymbol{\Phi} \cdot \begin{pmatrix} (\rho_x)_{ss'} \\ \tau(\rho_y)_{ss'} \end{pmatrix} a^\dagger_{\sigma,\ell,s,\tau}(\boldsymbol{r}) a_{\sigma,\ell,s',\tau}(\boldsymbol{r})$$

where we have made the reasonable assumption that the order parameter is trivial in spin space and does not possess any additional layer-hopping or layer-dependent component. Note that $\rho_j$, in the above expression, stands for Pauli matrices in the sublattice space. In Fig. 4 of the main text, we set $\Phi_0 = 20$ meV for the graphene nematic order parameter; note that even with this somewhat large value, the changes in the LDOS with respect to the non-nematic phase are nearly imperceptible.

**Supplementary References**